\documentstyle{aipproc}
\begin{document}
\title{Three-dimensional topological\\
quantum field theory of Witten type%
\thanks{Talk delivered by B.~B.} }
\author{Ma\l gorzata Bakalarska$^*$ and Bogus\l aw Broda$^{\dagger}$}  
\address{Department of Theoretical Physics\\
University of \L\'od\'z, Pomorska 149/153\\
PL--90-236 \L\'od\'z, Poland\\
\tt
$^*$gosiabak@krysia.uni.lodz.pl\\
$^{\dagger}$bobroda@krysia.uni.lodz.pl}
\maketitle
\begin{abstract}
Description of two three-dimensional topological quantum field theories of Witten type as twisted supersymmetric theories is presented. Low-energy effective action and a corresponding topological invariant of three-dimensional manifolds are considered.
\end{abstract}

\section{Introduction}
The aim of our work is to present a step made in the direction of understanding of three-dimensional (3d) topological quantum field theory (TQFT) in the spirit Witten and Seiberg have done in the four-dimensional case. Actually, we will consider two ``microscopic'' 3d TQFT's and sketch their common low-energy consequences for topology of 3d manifolds.

First of all, let us recall the milestones in the development of four-dimensional TQFT.
In 1988, Witten proposed a description of Donaldson's theory (``topological'' invariants of four-dimensional manifolds) in terms of an appropriately twisted ${\cal N}=2$ SUSY SU(2) pure gauge theory \cite{w1}. But not so many mathematical consequences had followed from this approach until 1994 when Seiberg-Witten (SW) theory entered the scene. It has appeared that a newly-discovered dual description of the ${\cal N}=2$ theory in low-energy limit \cite{sw} provides us a new, alternative (but essentially equivalent and simpler) formulation of ``topological'' invariants of four-dimensional manifolds \cite{w2}.

In the first of our previous papers \cite{b}, a physical scenario has been proposed to reach our present goal using a geometric-topological construction with scalar curvature distribution ``compatible'' with surgery. Though the scalar curvature distribution used ``agrees'' with the surgery procedure, nevertheless it is unclear why such a distribution should be privileged. In the second paper \cite{bb}, we have proposed a simpler and more natural mechanism without any reference to curvature. In fact, we have directly applied known results concerning low-energy limit of 3d ${\cal N}=4$ SUSY gauge theory \cite{isw}.

\section{Topological field theory}
In 3d, we have the two important topological quantum field theories (of ``cohomological'' type): topological SU(2) gauge theory of flat connection and 3d version of the (topological) SW theory. The former is a 3d twisted ${\cal N}=4$ SUSY SU(2) pure gauge theory or a 3d version of the Donaldson-Witten (DW) theory, and ``by definition'' it describes the Casson invariant that appropriately counts the number of flat SU(2) connections \cite{bt} (see, Section A). The latter (3d SW) is a 3d twisted version of ${\cal N}=4$ SUSY U(1) gauge theory with a matter hypermultiplet \cite{sw}, \cite{w2} (see, Section B).
It is interesting to note that the both theories can be derived from 4d ${\cal N}=2$ SUSY SU(2) pure gauge theory corresponding via twist to DW theory.

\subsection{Donaldson-Witten and Casson theory}
Let us consider first the usual ${\cal N}=2$ supersymmetric SU(2)
Yang-Mills theory in flat Euclidean space $R^{4}$ \cite{ws}. The Yang-Mills
gauge field $A_{\mu}$ is embedded in the ${\cal N}=2$ chiral
multiplet ${\cal A}$ consisting of one ${\cal N}=1$ chiral multiplet
$\Phi = (B, {\psi}_{\alpha})$ and one ${\cal N}=1$
vector multiplet $W_{\alpha} = ({\lambda}_{\alpha}, A_{\mu})$.%
\footnote{We will use the conventions of Wess and Bagger \cite{ws}. For
  instance, doublets of the ${\rm SU(2)}_{L}$ (or ${\rm SU(2)}_{R}$) rotation
  symmetries are represented by spinor indices $\alpha,\beta,\ldots =
  1,2$ (or $\dot{\alpha},\dot{\beta},\ldots = 1,2$). Doublets of the
  internal ${\rm SU(2)}_{I}$ symmetry will be denoted by indices $i,j,
  \ldots = 1,2$. These indices are raised and lowered with the
  antisymmetric tensor ${\varepsilon}_{\alpha\beta}$ (or
  ${\varepsilon}_{\dot{\alpha}\dot{\beta}}$, ${\varepsilon}_{ij}$)
  with sign convention such that ${\varepsilon}_{12} = 1 =
  {\varepsilon} ^{21}$. Tangent vector indices are
  denoted as $\mu,\nu, \ldots = 1,\ldots,4$. The spinor and tangent
  vector indices are related with the tensors
  ${\sigma}_{\mu\alpha\dot{\beta}} = (-1,-i {\tau}^{a}),
  {\bar{\sigma}}^{\dot{\alpha}\beta}_{\mu} = (-1,i {\tau}^{a})$
  described in the Appendices A and B of
\cite{ws}
 by letting ${\eta}
  _{\mu\nu} \rightarrow - {\delta}_{\mu\nu}$ and ${\varepsilon}^{\mu
    \nu\rho\sigma} \rightarrow -i {\varepsilon}_{\mu\nu\rho\sigma}$
  with ${\varepsilon}_{1234} = 1$. Covariant derivatives are defined
  by $D_{\mu} B = \partial_\mu B - [A_{\mu},B]$, and the Yang-Mills
  field strength is $F_{\mu\nu} = \partial_\mu A_{\nu} - \partial_\nu
  A_{\mu} - [A_{\mu},A_{\nu}]$.}
The ${\cal N}=2$ chiral multiplet can be arranged in a diamond form,
\begin{eqnarray*}
   & A^{a}_{\mu} & \\
  {\lambda}^{a}_{\alpha} & &{\psi}^{a}_{\alpha},  \\
   & B^{a} &
\end{eqnarray*} 
to exhibit the ${SU(2)}_{I}$ symmetry which acts on the rows. This
theory is described by the action
\begin{eqnarray}
  S & = & \int_{R^{4}} d^{4} x {\rm Tr} \left( \frac{1}{4} F_{\mu\nu}
  F_{\mu\nu} + D_{\mu} \bar{B} D_{\mu} B + \frac{1}{2} [B,\bar{B}]^{2}
  - {\bar{\lambda}}_{\dot{\alpha} i} {\bar{\sigma}}^{\dot{\alpha}
  \alpha}_{\mu} D_{\mu} {{\lambda}_{\alpha}}^{i} + \mbox{}\right.
  \nonumber \\
  &&\left. \mbox{} - \frac{i}{\sqrt{2}} \bar{B} \varepsilon_{ij}
  [\lambda^{i}, \lambda ^{j}] + \frac{i}{\sqrt{2}} B
  {\varepsilon}^{ij} [{\bar{\lambda}}_{i},{\bar{\lambda}}_{j}]\right),
\label{2.1}
\end{eqnarray}
where ${{\lambda}_{\alpha}}^{i} = (\lambda_{\alpha},\psi_{\alpha})$.

Now we will construct twisted TQFT.
The rotation group $K$ in four-dimensional Euclidean space is locally
${\rm SU(2)}_{L} \times {\rm SU(2)}_{R}$. In addition, the connected component
of the global symmetry group of the ${\cal N}=2$ theory is ${\rm SU(2)}_{I}$. The
theory, when formulated on a flat $R^{4}$, therefore has a global
symmetry group \cite{w4}

\[
H = {\rm SU(2)}_{L} \times {\rm SU(2)}_{R} \times {\rm SU(2)}_{I}.
\]
Let ${\rm SU(2)}_{R^{\prime}}$ be a diagonal subgroup of ${\rm SU(2)}_{R}
\times {\rm SU(2)}_{I}$ obtained by sending ${\rm SU(2)}_{I}$ index ``$i$'' to
dotted index ``$\dot{\alpha}$'', and let
\[
K^{\prime} = {\rm SU(2)}_{L} \times {\rm SU(2)}_{R^{\prime}}.
\]
Then, the transformations of the fields under ${\rm SU(2)}_{L} 
\times {\rm SU(2)}_{R^{\prime}}$ are
\begin{eqnarray*}
  A_{\alpha\dot{\alpha}}: \left(\frac{1}{2},\frac{1}{2},0\right) &
  \longrightarrow & A_{\alpha\dot{\alpha}}: \left(\frac{1}{2},
    \frac{1}{2}\right), \\
  {\lambda}_{\alpha i}: \left(\frac{1}{2},0,\frac{1}{2}\right) &
  \longrightarrow & {\lambda}_{\alpha\dot{\beta}}: \left(\frac{1}{2},
    \frac{1}{2}\right), \\
  {\bar{\lambda}}_{\dot{\alpha} i}: \left(0, \frac{1}{2},\frac{1}{2}
  \right) & \longrightarrow &
  {\bar{\lambda}}_{\dot{\alpha}\dot{\beta}}: (0,0) \oplus (0,1), \\
  B: (0,0,0) & \longrightarrow & B: (0,0).
\end{eqnarray*} 
And, we decompose the gaugino doublet ${{\lambda}_{\alpha}}^{i}$ into
 $K^{\prime}$ irreducible representations as \cite{y}:
\begin{eqnarray}
  {\lambda}_{\alpha\dot{\beta}} & = & \frac{1}{\sqrt{2}} {\sigma}_{\mu
    \alpha\dot{\beta}} {\psi}_{\mu}, \nonumber \\
  {\bar{\lambda}}_{\dot{\alpha}\dot{\beta}} & = & \frac{1}{\sqrt{2}}
  \left({\bar{\sigma}}_{\mu\nu\dot{\alpha}\dot{\beta}} {\chi}_{\mu\nu}
  + \varepsilon_{\dot{\alpha}\dot{\beta}}\eta \right).
\label{2.2}
\end{eqnarray}
Substituting (\ref{2.2}) into the Lagrangian (\ref{2.1}), we have
\begin{equation}
  S = \int_{R^{4}} d^{4} x \left({\cal L}_{1} + {\cal L}_{2} +
      {\cal L}_{3} \right), 
\label{2.3}
\end{equation} 
where
\begin{eqnarray*}
&& {\cal L}_{1} = {\rm Tr} \left( \frac{1}{4} F_{\mu\nu} F_{\mu\nu}
   - {\chi}_{\mu\nu} \left(D_{\mu} {\psi}_{\nu} - D_{\nu} {\psi}_{\mu}
   \right) + \frac{i}{\sqrt{2}} B \left\{ {\chi}_{\mu\nu},{\chi}_{\mu\nu}
  \right\} \right),\\
&& {\cal L}_{2} = {\rm Tr} \left( - \eta D_{\mu} {\psi}_{\mu} + D_{\mu}
   \bar{B} D_{\mu} B - \frac{i}{\sqrt{2}} \bar{B} \left\{ {\psi}_{\mu},
   {\psi}_{\mu} \right\} \right), \\
&& {\cal L}_{3} = {\rm Tr} \left( \frac{1}{2} [B,\bar{B}]^{2} + 
   \frac{i}{\sqrt{2}} B \left\{ \eta,\eta \right\} \right).
\end{eqnarray*}
Thus, we have obtained a TQFT --- Donaldson-Witten
theory.

Now we simply assume all the fields to be independent of the fourth
coordinate and discard all mention of the fourth coordinate from the
Lagrangian. The dimensionally reduced version of (\ref{2.3}) is
\begin{eqnarray}
  S & = & \int_{R^3} d^3 x {\rm Tr} \left\{ \frac{1}{4} F_{mn} F_{mn} +
    \frac{1}{2} D_{m} \varphi D_{m} \varphi + 2 \chi_{m}
    \left[\varphi, \psi_{m}\right] - 2 \omega D_{m} {\chi}_{m} +
    \mbox{}\right.
  \nonumber \\
  && \mbox{} - 2 {\varepsilon}_{mnk} {\chi}_{k} D_{m} {\psi}_{n} +
  \frac{4i}{\sqrt{2}} B \left\{ {\chi}_{m},{\chi}_{m} \right\} +
  \eta [\phi,\omega] - \eta D_{m} {\psi}_{m} + \mbox{} \label{2.4} \\
  && \mbox{} + D_{m} \bar{B} D_{m} B + [\varphi,\bar{B}] [\varphi,B]
  - \frac{i}{\sqrt{2}} \bar{B} \left\{ \psi_{m},\psi_{m} \right\} -  
  \frac{i}{\sqrt{2}} \bar{B} \left\{ \omega, \omega \right\} +
  \mbox{} \nonumber \\
  &&\left. \mbox{} + \frac{1}{2} [B,\bar{B}]^{2} + \frac{i}{\sqrt{2}} B
    \left\{ \eta,\eta \right\}\right\}.\nonumber
\end{eqnarray}

We can obtain a supersymmetric action in three dimensions from
(\ref{2.1}) by the process of dimensional reduction,
\begin{eqnarray}
  S & = & \int_{R^3} d^3 x {\rm Tr} \left\{ \frac{1}{4} F_{mn} F_{mn} +
    \frac{1}{2} D_{m} \varphi D_{m} \varphi + D_{m} \bar{B} D_{m} B +
    \mbox{}\right. \nonumber \\
  && \mbox{} + [\varphi,\bar{B}] [\varphi,B] + \frac{1}{2}
  [B,\bar{B}]^{2} - i {\bar{\lambda}}_{\alpha\beta}
  {\sigma}^{\alpha\gamma}_{m}
  D_{m} {{\lambda}_{\gamma}}^{\beta} + \mbox{} \label{2.5} \\
  &&\left. \mbox{} + {\bar{\lambda}}_{\alpha\beta} [\varphi,{\lambda}^
    {\beta\alpha}] - \frac{i}{\sqrt{2}} \bar{B} {\varepsilon}_{ij}
    [{\lambda}^{i},{\lambda}^{j}] + \frac{i}{\sqrt{2}} B
    {\varepsilon}^{ij} [{\bar{\lambda}}_{i},{\bar{\lambda}}_{j}]
  \right\}. \nonumber
\end{eqnarray}
And then substituting the above-mentioned twist
\begin{eqnarray}
  {\lambda}_{\alpha\beta} & = & \frac{1}{\sqrt{2}} \left({\sigma}_
    {4\alpha\beta} \omega - i {\sigma}_{m\alpha\beta} {\psi}_{m}
    \right), \nonumber \\
  {\bar{\lambda}}_{\alpha\beta} & = & \frac{1}{\sqrt{2}} \left(2i
    {\sigma}_{k\alpha\beta} {\chi}_{k} +
    {\varepsilon}_{\alpha\beta}\eta \right),
\label{2.6}
\end{eqnarray}
into the action (\ref{2.5}) we get the same action (\ref{2.4}).

Dropping fermions and topologically trivial bosons in equation (\ref{2.5}) we obtain
\[
S = \int_{R^{3}} d^{3} x {\rm Tr} \left( \frac{1}{4} F_{mn}
    F_{mn} \right).
\]
It is a part of the bosonic action, with the absolute minima equation,
\[
F_{mn}^a = 0,
\]
corresponding to the Casson invariant of 3d manifolds \cite{bt}.

\subsection{Seiberg-Witten theory}
We start again from the ${\cal N}=2$ SUSY SU(2) Yang-Mills theory. We recall that the moduli space of the ${\cal N}=2$ SUSY
SU(2) Yang-Mills theory \cite{sw} contains two singular points.
At these points the low energy effective theory is ${\cal N}=2$ SUSY
U(1) theory coupled to an additional massless matter (monopoles or
dyons) in the form of the ${\cal N}=2$ hypermultiplet (sometimes
called the scalar multiplet):%
\footnote{Here the low energy fields are: the vector
multiplet (gauge field $A_{\mu}$, ${\rm SU(2)}_{I}$ doublet of fermions
${{\lambda}_{\alpha}}$ and ${\psi}_{\alpha}$, a complex scalar $B$)
and the hypermultiplet (consisting of two Weyl fermions ${\psi}_{A}$
and $\psi^{\dag}_{\tilde{A}}$ and complex bosons $A$ and
$\tilde{A}^{\dag}$).}
\begin{eqnarray*}
     & {\psi}_{A} & \\
          A~ & & \tilde{A}^{\dag}. \\
     & \psi^{\dag}_{\tilde{A}} &
\end{eqnarray*}
The action of an ${\cal N}=2$ supersymmetric abelian gauge theory
coupled to a massless hypermultiplet is given by
\begin{eqnarray}
  S & = & \int_{R^4} d^4 x \left\{\frac{1}{4} F_{\mu\nu}
  F_{\mu\nu} + \partial_\mu \bar{B} \partial_\mu B - 
  {\bar{\lambda}}_{\dot{\alpha}i}
  {\bar{\sigma}}^{\dot{\alpha}\alpha}_{\mu} \partial_\mu
  {{\lambda}_{\alpha}}^{i} + D_{\mu} {\bar{A}}^{i} D_{\mu}
  A_{i} + \mbox{}\right. \nonumber \\
  && \mbox{} + \frac{1}{2} ({\bar{A}}^{i} A_{i})^{2} + i \bar{\psi}
  {\gamma}^{\mu} D_{\mu} \psi + i \sqrt{2} {\bar{A}}^{i}
  {\bar{\lambda}}_{i} \psi - i \sqrt{2} \bar{\psi} {\lambda}^{i} A_{i}
  - \bar{\psi} (B - {\gamma}_{5} \bar{B}) \psi + \mbox{} \label{2.7} \\
  &&\left. \mbox{} - {\bar{A}}^{i} (B^{2} + {\bar{B}}^{2}) A_{i}
  \right\}. \nonumber
\end{eqnarray}

 We know that the twist consists of considering as the rotation
 group the group, $K^{\prime} = {\rm SU(2)}_{L} \times {\rm SU(2)}_{R^
 {\prime}}$ and this implies that the hypermultiplet field content is modified 
 as follows:
\begin{eqnarray*}
&& A^{i}: \left( 0, 0, \frac{1}{2} \right) \longrightarrow
   M^{\dot{\alpha}}: \left(\frac{1}{2}, 0 \right),\\
&& {\psi}_{A\alpha}: \left( \frac{1}{2}, 0, 0 \right) \longrightarrow
   u_{\alpha}: \left( \frac{1}{2}, 0 \right),\\
&& {\bar{\psi}}_{\tilde{A}\dot{\alpha}}: \left( 0, \frac{1}{2}, 0 
   \right) \longrightarrow {\bar{v}}_{\dot{\alpha}}: \left( 0, 
   \frac{1}{2}\right),\\
&& A^{\dagger}_{i}: \left( 0, 0, \frac{1}{2} \right) \longrightarrow
   {\bar{M}}_{\dot{\alpha}}: \left( \frac{1}{2}, 0 \right),\\
&& {\bar{\psi}}_{A\dot{\alpha}}: \left( 0, \frac{1}{2}, 0 \right) 
   \longrightarrow {\bar{u}}_{\dot{\alpha}}: \left( 0, \frac{1}{2}
   \right),\\
&& {\psi}_{\tilde{A}\alpha}: \left( \frac{1}{2}, 0, 0 \right) 
   \longrightarrow v_{\alpha}: \left( \frac{1}{2}, 0 \right).
\end{eqnarray*}
 Substituting  equation (\ref{2.2}) into the action (\ref{2.7}) 
 and taking into account that field content we get the following
 twisted euclidean action (compare to \cite{ll})
\begin{eqnarray}
  S & = & \int_{R^4} d^4 x \left\{ \frac{1}{4} F_{\mu\nu}
  F_{\mu\nu} + \partial_\mu \bar{B} \partial_\mu B - 2 \chi_{\mu\nu}
  \partial_\mu {\psi}_{\nu} - \eta \partial_\mu \psi_{\mu} + D_{\mu}
  {\bar{M}}^{\dot{\alpha}} D_{\mu} M_{\dot{\alpha}} +
  \mbox{} \right. \nonumber \\
  && \mbox{} + \frac{1}{2} ({\bar{M}}^{\dot{\alpha}}
  M_{\dot{\alpha}})^{2} - i \left({\bar{M}}^{\dot{\beta}} {\psi}_
  {\alpha\dot{\beta}} u^{\alpha} - v^{\alpha} {\psi}_{\alpha
  \dot{\beta}} M^{\dot{\beta}}\right) + i \eta
  \left({\bar{u}}^{\dot{\alpha}} M_{\dot{\alpha}} - {\bar{M}}^{\dot
  {\beta}} {\bar{v}}_{\dot{\beta}}\right) + \mbox{} \label{2.8} \\
  && \mbox{} + i {\chi}_{\dot{\alpha} \dot{\beta}} ({\bar{M}}^
  {\dot{\alpha}} {\bar{v}}^{\dot{\beta}} - {\bar{u}}^{\dot{\alpha}}
  M^{\dot{\beta}}) - B^{2} {\bar{M}}^{\dot{\alpha}} M_{\dot{\alpha}} -
  {\bar{B}}^{2} {\bar{M}}^{\dot{\alpha}} M_{\dot{\alpha}} + v^{\alpha}
  \bar{B} u_{\alpha} + \mbox{} \nonumber \\
  &&\left. \mbox{} - B v^{\alpha} u_{\alpha} - B
  {\bar{u}}_{\dot{\alpha}} {\bar{v}}^{\dot{\alpha}} - \bar{B}
  {\bar{u}}_{\dot{\alpha}} {\bar{v}} ^{\dot{\alpha}} + i v^{\alpha}
  D_{\alpha \dot{\alpha}} {\bar{v}} ^{\dot{\alpha}} + i
  {\bar{u}}^{\dot{\alpha}} D_{\alpha \dot{\alpha}} u^{\alpha}
  \right\}.\nonumber
\end{eqnarray}
The dimensionally reduced version of (\ref{2.8}) is
\begin{eqnarray}
  S & = & \int_{R^3} d^3 x \left\{ \frac{1}{4} F_{mn} F_{mn}
  + \frac{1}{2} \partial_m \varphi \partial_m \varphi + \partial_m
  \bar{B} \partial_m B - 2 {\varepsilon}_{mnk} {\chi}_{k}
  \partial_m {\psi}_{n} + \mbox{}\right. \nonumber \\
  && \mbox{} - 2 \omega \partial_m \chi_{m} - \eta \partial_{m}
  {\psi}_{m} + D_{m} {\bar{M}}^{\alpha} D_{m} M_{\alpha} + [\varphi,
  {\bar{M}}^{\alpha}] [\varphi,M_{\alpha}] + \mbox{} \nonumber \\
  && \mbox{} + \frac{1}{2} ({\bar{M}}^{\alpha} M_{\alpha})^{2} + i
  \omega \left({\bar{M}}^{\beta} u_{\beta} - v^{\alpha}
  M_{\alpha}\right) + v^{\alpha} {\psi}_{\alpha\beta} M^{\beta} -
 {\bar{M}}^{\beta} {\psi}_{\alpha\beta} u^{\alpha} + \mbox{}\label{2.9}\\
  && \mbox{} + i \eta \left({\bar{u}}^{\alpha} M_{\alpha} -
  {\bar{M}}^{\beta} {\bar{v}}_{\beta}\right) + 2
  {\chi}_{\alpha\beta} \left({\bar{u}}^{\alpha} M^{\beta} -
  {\bar{M}}^{\alpha} {\bar{v}}^{\beta}\right) - B^{2} {\bar{M}}
  ^{\alpha} M_{\alpha} + \mbox{} \nonumber \\
  && \mbox{} - {\bar{B}}^{2} {\bar{M}}^{\alpha} M_{\alpha} + \bar{B}
  v^{\alpha} u_{\alpha} - B v^{\alpha} u_{\alpha} - B
  {\bar{u}}_{\alpha} {\bar{v}}^{\alpha} - \bar{B} {\bar{u}}_{\alpha}
  {\bar{v}}^{\alpha} + v^{\alpha} D_{\alpha \beta} {\bar{v}}^{\beta}
  + \mbox{} \nonumber \\
  &&\left. \mbox{} + {\bar{u}}^{\beta} D_{\alpha\beta} u^{\alpha} + i
  v^{\alpha} [\varphi,{\bar{v}}_{\alpha}] + i {\bar{u}}^{\alpha}
  [\varphi,u_{\alpha}] \right\}.\nonumber
\end{eqnarray}

As a result of the dimensional reduction of (\ref{2.7}) we get
\begin{eqnarray}
  S & = & \int_{R^3} d^3 x \left\{ \frac{1}{4} F_{mn}
  F_{mn} + \frac{1}{2} \partial_m \varphi \partial_m \varphi +
  \partial_m  \bar{B} \partial_m B - i {\bar{\lambda}}_{\alpha i}
  {\sigma}^{\alpha\beta}_{m} \partial_m {{\lambda}_{\beta}}^{i} +
    \mbox{}\right.\nonumber \\
  && \mbox{} + D_{m} {\bar{A}}^{i} D_{m} A_{i} + [\varphi,{\bar{A}}^{i}]
  [\varphi,A_{i}] + \frac{1}{2} ({\bar{A}}^{i} A_{i})^{2} -
  {\bar{A}}^{i} (B^{2} +  {\bar{B}}^{2}) A_{i} + \mbox{} \nonumber \\
  && \mbox{} + i \sqrt{2} {\bar{A}}^{i} {\bar{\lambda}}_{\alpha i}
  {\bar{\psi}}^{\alpha}_{\tilde{A}} - i \sqrt{2} {\bar{\psi}}_{A 
  \alpha} {\bar{\lambda}}^{\alpha i} A_{i} - i \sqrt{2} {\bar{A}}^{i}
  {\lambda}_{\alpha i} {\psi}^{\alpha}_{A} + i \sqrt{2} {\psi}^
  {\alpha}_{\tilde{A}} {\lambda}_{\alpha i} A^{i} + \mbox{}\label{2.10}\\
  && \mbox{} + \bar{B} {\psi}^{\alpha}_{\tilde{A}} {\psi}_{A\alpha}
  - B {\psi}^{\alpha}_{\tilde{A}} {\psi}_{A\alpha} - B {\bar{\psi}}_
  {A\alpha} {\bar{\psi}}^{\alpha}_{\tilde{A}} - \bar{B} {\bar{\psi}}_
  {A\alpha} {\bar{\psi}}^{\alpha}_{\tilde{a}} + {\psi}^{\alpha}_
  {\tilde{A}} D_{\alpha\beta} {\bar{\psi}}^{\beta}_{\tilde{A}} +
  \mbox{} \nonumber \\
  &&\left. \mbox{} + {\bar{\psi}}^{\beta}_{A} D_{\alpha\beta} 
  {\psi}^{\alpha}_{A} + i {\psi}^{\alpha}_{\tilde{A}} [\varphi, 
  {\bar{\psi}}_{\tilde{A}\alpha}] + i {\bar{\psi}}^{\alpha}_{A} 
  [\varphi, {\psi}_{A\alpha}] \right\}. \nonumber
\end{eqnarray}

Using the twist
\begin{eqnarray*}
&& {\lambda}_{\alpha\beta} = \frac{1}{\sqrt{2}} \left( {\sigma}_{4
 \alpha\beta} \omega - i {\sigma}_{m\alpha\beta} {\psi}_{m} \right),\\
&& {\bar{\lambda}}_{\alpha\beta} = \frac{1}{\sqrt{2}} \left( 
  2 i {\sigma}_{k\alpha\beta} {\chi}_{k} + {\varepsilon}_{\alpha\beta}
  \eta \right),\\
&& A^{i} \rightarrow M^{\alpha}, \\
&& {\psi}_{A\alpha} \rightarrow u_{\alpha}, \\
&& {\bar{\psi}}_{\tilde{A} \alpha} \rightarrow {\bar{v}}_{\alpha},\\
&& {\bar{A}}_{i} \rightarrow {\bar{M}}_{\alpha},\\
&& {\bar{\psi}}_{A\alpha} \rightarrow {\bar{u}}_{\alpha},\\
&& {\psi}_{\tilde{A} \alpha} \rightarrow v_{\alpha},
\end{eqnarray*}
 we get the action (\ref{2.9}).

Dropping fermions and topologically trivial bosons we obtain
\begin{equation}
 S  = \int_{R^3} d^3 x
\left(\frac{1}{4} F_{mn} F_{mn} + \frac{1}{2}
           ({\bar{A}}^{i} A_{i})^{2} + D_{m} {\bar{A}}^{i}
            D_{m} A_{i}\right).
\label{2.11}
\end{equation}
It is a part of the bosonic matter action.

\section{low-energy consequences}
Interestingly, it follows from \cite{isw} that low-energy effective theory for both 3d ${\cal N}=4$ SUSY SU(2) case as well as for the 3d ${\cal N}=4$ SUSY abelian one with matter hypermultiplet is pure gauge abelian, and the moduli spaces are smooth.
More precisely, we have the so-called Atiyah-Hitchin manifold (interpreted as the two-monopole moduli space), a complete hyper-K\"ahler manifold, in the first case, and the Taub-NUT manifold, in the second one.

3d ${\cal N} = 4$ supersymmetric abelian gauge theory
is described by the following action%
\footnote{$\vec{G}$ is an auxiliary field.}
\begin{eqnarray}
 S & = & \int_{R^{3}} d^3 x \left( \frac{1}{4} F_{mn} F_{mn} +
   \frac{1}{2} \partial_m \varphi \partial_m \varphi + \partial_m
   \bar{B} \partial_m B + \mbox{}\right. \nonumber \\
   &&\left. \mbox{} - i {\bar{\lambda}}_{\alpha i} {\sigma}^{\alpha
   \beta}_{m} \partial_m {{\lambda}_{\beta}}^{i} - \frac{1}{2}
   (\vec{G})^{2} \right).
\label{3.1}
\end{eqnarray}
Introducing infinitesimal parameters ${{\xi}_{\alpha}}^{i}$ and
${\bar{\xi}}_{\alpha i}$, the supersymmetry transformations are given
by
\begin{eqnarray}
  && \delta A_{m}  =  i {\bar{\lambda}}_{\alpha i} {\sigma}^{\alpha
  \beta} _{m} {{\xi}_{\beta}}^{i} - i {\bar{\xi}}_{\alpha i}
  {\sigma}^{\alpha\beta}_{m} {{\lambda}_{\beta}}^{i}, \nonumber \\
  && \delta \varphi  =  {{\bar{\xi}}^{\alpha}}_{i} {{\lambda}_
  {\alpha}}^{i} - {\bar{\lambda}}_{\alpha i} {\xi}^{\alpha i},
  \nonumber \\
  && \delta B  =  - \sqrt{2} {\varepsilon}_{ij} {\xi}^{\alpha i}
  {{\lambda}_{\alpha}}^{j}, \nonumber \\
  && \delta \bar{B}  =  - \sqrt{2} {\varepsilon}^{ij}
  {{\bar{\xi}}^{\alpha}}_{i} {\bar{\lambda}}_{\alpha j},\label{3.2}\\
  && \delta {{\lambda}_{\alpha}}^{i}  =  - \frac{i}{2}
  {\varepsilon}_{mnk} {{\sigma}_{k \alpha}}^{\beta} {{\xi}_{\beta}}^{i}
  F_{mn} - i \sqrt{2} {\varepsilon}^{ij} {\sigma}_{m\alpha\beta}
  {{\bar{\xi}}^{\beta}}_{j} \partial_m B + \mbox{}  \nonumber \\
  && \mbox{} - i {{\sigma}_{m \alpha}}^{\beta} {{\xi}_{\beta}}^{i}
  \partial_m \varphi, \nonumber \\
  && \delta {\bar{\lambda}}_{\alpha i}  =  i \sqrt{2}
  {\varepsilon}_{ij} {\sigma}_{m\alpha\beta} {\xi}^{\beta j}
  \partial_m \bar{B} + \frac{i}{2} {\varepsilon}_{mnk}
  {{\sigma}_{k\alpha}}^{\beta}
  {\bar{\xi}}_{\beta i} F_{mn} + \mbox{} \nonumber \\
  && \mbox{} - i {{\sigma}_{m\alpha}}^{\beta} {\bar{\xi}}_{\beta i}
  \partial_m \varphi + 2 i {{\sigma}_{m\alpha}}^{\beta} {\bar{\xi}}
  _{\beta i} G_{m}, \nonumber \\
  && \delta \vec{G}  = i {\bar{\xi}}_{\alpha i} {{\vec{\tau}}^{i}}_{j}
  {\sigma}^{\alpha\beta}_{m} \partial_m {{\lambda}_{\beta}}^{j}.
  \nonumber
\end{eqnarray}
Using the equations of motion for all the fields one can  easily show 
that this action is invariant under the supersymmetric transformations (\ref{3.2}).
Taking into account the following substitution 
\begin{eqnarray*}
 {\lambda}_{\alpha\beta} = \frac{1}{\sqrt{2}} \left( {\sigma}_
 {4\alpha\beta} \omega - i {\sigma}_{m\alpha\beta} {\psi}_{m} 
 \right),\\
 {\bar{\lambda}}_{\alpha\beta} = \frac{1}{\sqrt{2}} \left( 2 i 
 {\sigma}_{k\alpha\beta} {\chi}_{k} + {\varepsilon}_{\alpha\beta}
 \eta \right),
\end{eqnarray*}
and knowing the auxiliary field $\vec{G}$ transforms in the
$(1,0)$ representation (and hence becomes a vector $H_{k}$), we get the
following 3d twisted version of the action (\ref{3.1})
\begin{eqnarray}
  S & = & \int_{R^3} d^3 x \left( \frac{1}{4} F_{mn} F_{mn} +
    \frac{1}{2} \partial_m \varphi \partial_m \varphi + \partial_m
    \bar{B} \partial_m B + 2 {\varepsilon}_{mnk} {\psi}_{n} \partial_m
    {\chi}_{k} + \mbox{}\right. \nonumber \\
  &&\left. \mbox{} - 2 \omega \partial_m {\chi}_{m} - \eta \partial_m
    {\psi}_{m} - 2 H_{k} H_{k} \right).
\label{3.3}
\end{eqnarray}
The topological BRST transformation ${\delta}_{\rm B}$ induced by the
${\cal N}=4$ supertransformations is found by putting $\xi = 0$, $\bar{\xi} =
\frac{{\varepsilon}_{\alpha\beta} \cdot \rho}{\sqrt{2}}$, which reads
$\delta = - \rho \cdot {\delta}_{\rm B}$.%
\footnote{$\rho$ is an
anticommuting parameter and the lower index B denotes BRST
transformation.}
The BRST transformations are
\begin{eqnarray}
&& {\delta}_{\rm B} A_{m}  =  {\psi}_{m}, \nonumber \\
&& {\delta}_{\rm B} \varphi  =  - \omega, \nonumber \\
&& {\delta}_{\rm B} B  =  {\delta}_{\rm B} \omega = {\delta}_{\rm B} \eta = 0,
  \nonumber \\
&& {\delta}_{\rm B} \bar{B}  =  - \sqrt{2} \eta, \label{3.4} \\
&& {\delta}_{\rm B} {\psi}_{m}  =  - \sqrt{2} \partial_m B,
  \nonumber \\
&& {\delta}_{\rm B} {\chi}_{k}  =  - \frac{1}{2} \partial_k \varphi -
  \frac{1}{2} {\varepsilon}_{kmn} \partial_m A_{n} - H_{k}, \nonumber \\
&& {\delta}_{\rm B} H_{k}  =  \frac{1}{2} \partial_k \omega - \frac{1}{2}
  {\varepsilon}_{kmn} \partial_m {\psi}_{n}. \nonumber
\end{eqnarray} 

These BRST transformations are off-shell nilpotent up to a gauge
transformation ${\delta}^{2}_{\rm B} = - \sqrt{2} {\delta}_{\rm gauge}$; e.g.,
${\delta}^{2}_{\rm B} A_{m} = - \sqrt{2} \partial_m B$.  The action
(\ref{3.3}) is also nilpotent but on-shell.

Confining ourselves to the contribution of the Coulomb branch coming from abelian flat connections on the 3d manifold
${\cal M}^3$ we should consider a mathematical object akin to the Casson invariant. We have argued in \cite{b} that such an object should count (algebraically) the number of abelian flat connection on a cover of ${\cal M}^3$. If ${\cal M}^3$ arises from 
${\cal S}^3$ via 0-framed surgery on a knot \cite{r} the 3d invariant is directly related to the Alexander invariant of ${\cal M}^3$ \cite{mthl}.

\section{Conclusions}
In this paper, we have indicated that both 3d TQFT of SU(2) flat connection as well as 3d version of topological SW theory can be described in low-energy regime by the Alexander invariant. In a future work, it will be necessary to improve the analysis to include all contribution coming from the Coulomb branch.

\section{Acknowledgments}
The work has been supported by KBN grant 2 P03B 084 15 and by University of \L\'od\'z\ grant 621.

\end{document}